\documentclass[pre,twocolumn,superscriptaddress,showpacs]{revtex4}
\usepackage{graphicx}
\usepackage{graphicx}
\usepackage{amssymb}
\usepackage{amsmath}

\def\be{\begin{equation}}
\def\ee{\end{equation}}
\def\BE{\begin{equation}}
\def\EE{\end{equation}}
\def\BA{\begin{eqnarray}}
\def\EA{\end{eqnarray}}

\begin{document}

\title{Anomalous scaling in an age-dependent branching model}

\author{Stephanie Keller-Schmidt}
\affiliation{Bioinformatics, Institute of Computer Science, University
Leipzig, H{\"a}rtelstr. 16-18, 04107 Leipzig, Germany}
\author{Murat Tu\u{g}rul}
\affiliation{IST Austria, Am Campus 1, 3400 Klosterneuburg, Austria}
\author{V{\'\i}ctor M. Egu{\'\i}luz}
\affiliation{IFISC (CSIC-UIB), Instituto de F\'isica
Interdisciplinar y Sistemas Complejos, E-07122 Palma de
Mallorca, Spain}
\author{Emilio Hern\'andez-Garc{\'\i}a}
\affiliation{IFISC (CSIC-UIB), Instituto de F\'isica
Interdisciplinar y Sistemas Complejos, E-07122 Palma de
Mallorca, Spain}
\author{Konstantin Klemm}
\affiliation{Bioinformatics, Institute of Computer Science, University
Leipzig, H{\"a}rtelstr. 16-18, 04107 Leipzig, Germany}
\affiliation{Bioinformatics and Computational Biology, University of Vienna,
W\"{a}hringerstra{\ss}e 29, 1090 Vienna, Austria}
\affiliation{Theoretical Chemistry, University of Vienna,
W\"{a}hringerstra{\ss}e 17, 1090 Vienna, Austria}
\affiliation{School of Science and Technology, Nazarbayev University,
Kabanbay Batyr Ave.\ 53, 010000 Astana, Kazakhstan}

\date{\today}

\begin{abstract}
We introduce a one-parametric family of tree growth models, in which
branching probabilities decrease with branch age $\tau$ as $\tau^{-\alpha}$.
Depending on the exponent $\alpha$, the scaling of tree depth with tree size $n$ displays
a transition between the logarithmic scaling of random trees and an algebraic growth.
At the transition ($\alpha=1$) tree depth grows as $(\log n)^2$. This anomalous
scaling is in good agreement with the trend observed in evolution of biological
species, thus providing a theoretical support for age-dependent speciation and
associating it to the occurrence of a critical point.
\end{abstract}

\pacs{
89.75.Hc 
89.75.Da 
89.75.Fb 
87.23.Kg 
}
\maketitle
\section{Introduction}

Tree structures appear in a variety of contexts ranging from
river networks \cite{RodriguezIturbe1997} and blood vessels
\cite{West1997} to directed polymers
\cite{HalpinHealy1995,Brunet2008} or computer file systems
\cite{Klemm2005,Klemm2006,Geipel2009}. Evolutionary histories
and genealogies are naturally represented as trees. Each
branching point represents an ancestral relationship in a
population or an event of diversification on sets of languages
\cite{Gray2003}, species
\cite{Derrida1999,Serva2005,Herrada2008,Herrada2011,Brunet2013}
or socio-cultural innovations \cite{Sood2010}.
Based on genetic information, modern computational biology has
inferred thousands of trees, so-called phylogenies
\cite{FELSENSTEINInferringPhylo2003}, depicting the
evolutionary relationships between sets of species, from
bacteria to mammals \cite{Sanderson1994}. The {\em shapes} of
the collected phylogenies and of related evolutionary trees
\cite{Whelan06} share statistical properties not observed in
trees generated by standard branching models
\cite{Harris1963,Simkin2011,Pinelis2003}. It has been a
long-standing and fundamental question in evolutionary biology
to identify which processes accurately describe the observed
tree shapes and thus may serve as models of biological
evolution \cite{BLUMWhichRandomProcess2006,Jones2011}.

A suitable starting point and null hypothesis is the Equal Rate
Markov (ERM) process, which assumes that species speciate at a
constant homogeneous rate, independently of previous events and
of other species present. More specifically, starting from a
single tip (root), at each discrete time step a tip $i$ is
chosen uniformly at random and two new tips are attached to
$i$, increasing the number of tips by 1. The procedure has a
direct interpretation for macroevolution as a sequence of
speciation events, where the chosen species $i$ is the latest
common ancestor of two new species. The resulting topology of
the growing tree, which is equivalent to the one produced by
the Yule model \cite{Yule1925}, tends to generate compact and
nearly {\em balanced} tree shapes. Balance refers to an even
distribution of the number of nodes in the subtrees
arising from the branches created at each speciation point.
However, when comparing with the shape of large collections of
observed phylogenetic trees available nowadays (e.g.
\cite{Sanderson1994,Whelan06,Pompei2012}), the ERM hypothesis
can be rejected, as most real phylogenetic trees are
significantly less balanced than those generated by the ERM and
Yule models
\cite{BLUMWhichRandomProcess2006,Pompei2012,Herrada2008,Jones2011}.

\begin{figure}
\includegraphics[width=\columnwidth]{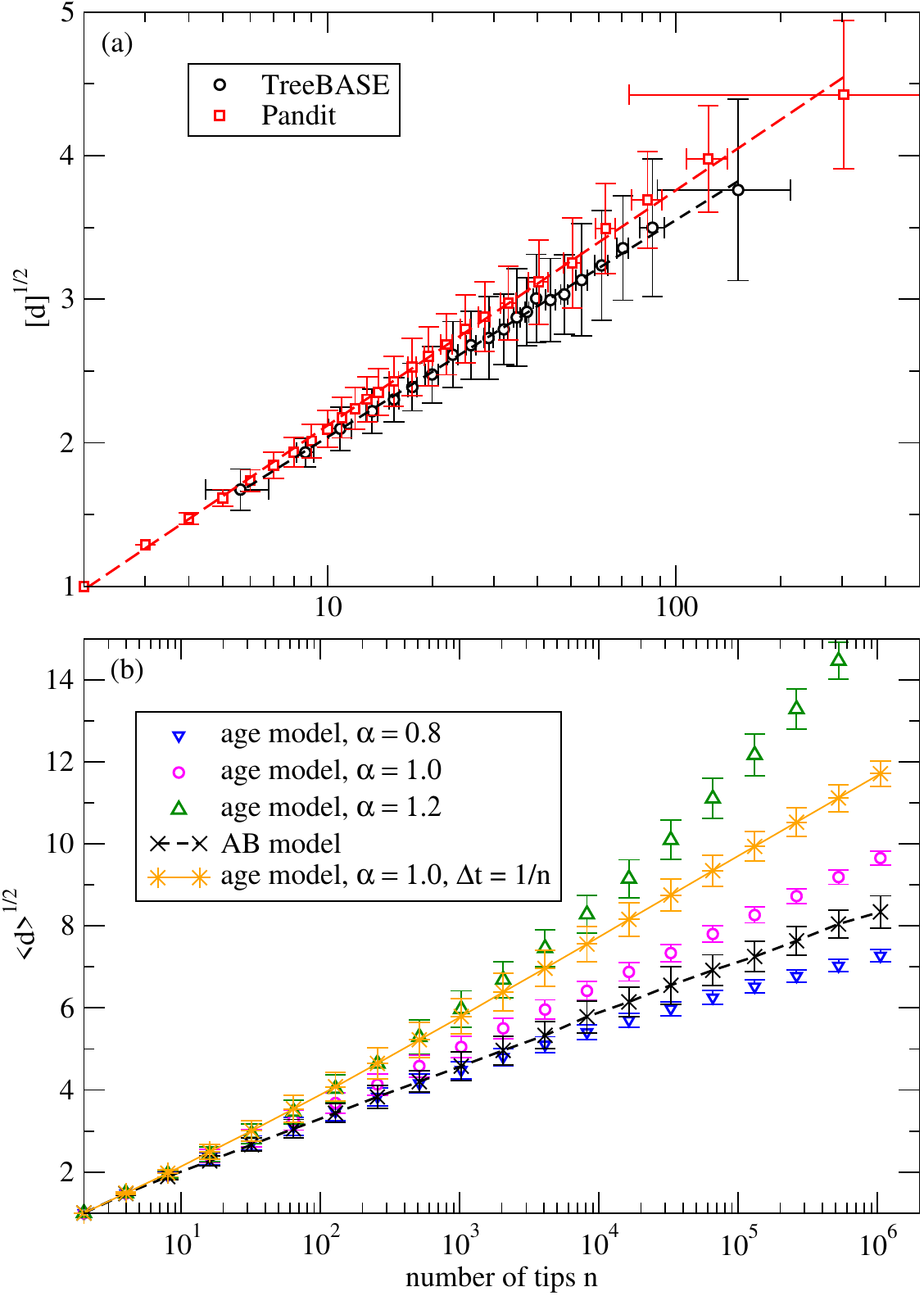}
\caption{ \label{fig:depth} (Color online)
Scaling of tree depth with size $n$. (a) Trees from databases TreeBASE and
PANDIT.
Trees have been binned by size such that each bin contains at least $s$ trees
using $s=1000$ for PANDIT and $s=200$ for TreeBASE. Least squares fits (dashed lines) of
the form
$\sqrt{d} = y = a x +b$ with $x =\ln n$ yield
$a =0.657 \pm 0.008$, $b=0.53 \pm 0.03$
  with correlation coefficient $r=0.9986$ for TreeBASE; and
$a =0.771 \pm 0.006$, $b=0.48 \pm 0.02$, $r=0.9990$ for PANDIT.
(b) Depth from {\sl age} and AB models. Fits analogous to the
above yield
$a=0.654 \pm 0.002$, $b=0.54 \pm 0.02$, $r=0.99995$ for the age model with
$\alpha =1.0$, $\Delta t=1$;
$a=0.556 \pm 0.003$, $b=0.72 \pm 0.02$, $r=0.998478$ for the AB model;
$a=0.822 \pm 0.006$, $b=0.20 \pm 0.05$, $r=0.99959$ for the age model with
$\alpha = 1.0$, $\Delta t = 1/n$. Vertical error bars indicate
(average $\pm$ standard deviation)$^{1/2}$, in (a) for the average $[d]$ taken over trees inside
a bin, in (b) for the mean depth $\langle d \rangle$ estimated by
100 independent realizations at each given size $n$. Horizontal error bars
in panel (a) give average $\pm$ std.\ dev.\ over the tree sizes inside
each bin.
}
\end{figure}

\section{Tree shape and depth}

Several indices for imbalance
measurement have been proposed, used and compared, see
Ref.~\cite{Pompei2012,MOOERSandHEARDReviewPhyloTreeShape1997,
MATSENGeometricApproachTreeShape2006,AGAPOWPURVISPowerOfEight2002}
for detailed discussion. Here we study how the depth
\cite{SACKINPhenogram1972} of a tree with $n$ tips
\begin{equation} \label{depth}
d = n^{-1} \sum_{i=1}^n d_i
\end{equation}
scales with $n$. For each tip $i$, $d_i$ denotes the number of
edges separating $i$ from the root. The role of the depth in
capturing tree imbalance is apparent by the two extreme cases.
For the (fully balanced) complete binary tree, $d = \log_2 n$
since all $n=2^k$ tips are at distance $k$ from root. On the
other extreme of full imbalance, a comb (or pectinate) tree has
$n$ tips attached to a path of $n-1$ nodes starting at the
root. Here $n d= \sum_{i=1}^n d_i= 1 + 2 + \dots + (n-2) + 2
(n-1)$, resulting in asymptotically linear scaling $d \sim n$.
For the ERM model, the small random imbalances introduced in
the process are not enough to affect the dominant scaling
behavior of the balanced tree and one finds $\langle d \rangle \sim
\log n$ (the average is over realizations of the random
process). This logarithmic scaling is a
robust outcome related to the exponential growth of tips
occurring in time for virtually any model of growing
supercritical trees \cite{Harris1963}, as far as branches split
independently and without memory, or if these correlations and
memory are sufficiently short-ranged. Therefore we denote
the logarithmic scaling of depth with tree size as {\em normal}.
Deviating scaling is called {\em anomalous}.

We have calculated \cite{Herrada2008,Herrada2011} the
depth $d$ for all trees (and subtrees) in the phylogenetic
databases TreeBASE (containing species phylogenies
\cite{Sanderson1994}) and PANDIT (protein phylogenies
\cite{Whelan06}). The result in Figure~\ref{fig:depth}(a)
suggests that the average depth grows with the number of tips
as
\begin{equation}
\langle d \rangle \sim (\log n)^2
\end{equation}
in good approximation. Although alternative scaling laws have
been proposed \cite{Herrada2008,Stich2009}, the $(\log n)^2$
form is more accurate for large tree sizes
\cite{Herrada2011,Keller-Schmidt2012}. Similar behavior is
observed in virus phylogenies where the scaling for individual
phylogenies was reported to follow the behavior $(\log
n)^\gamma$ with $\gamma$ varying from 1 to 3 \cite{Pompei2012}.
The important point is the departure from the $\log n$ scaling
of the ERM class. Thus strong correlations are important in the
evolutionary processes represented in the phylogenetic
databases.

\section{Statistical ensembles of trees}

A direct approach to capturing the imbalance of phylogenetic
trees is by defining a probability $\pi(l|n)$ of placing
exactly $l$ out of $n$ given tips in one of the two subtrees.
This stochastic splitting is first applied at the root and then
iterated at the roots of its two subtrees, at their subtrees'
roots and so forth, until arriving at the tips. A statistical
ensemble of trees is constructed by considering all possible
binary trees up to a given size, and assigning a probability to
each of them as just the product of the splitting probabilities
of all the inner nodes.

Choosing uniform probabilities independently of $l$, $\pi_{\rm
ERM} (l|n) = 1/(n-1)$ , $1\le l \le n-1$, leads to the ERM.
Aldous' Branching (AB) model \cite{ALDOUSBetaSplitting1996} is
the specific choice $\pi_\text{AB}(l|n) \propto
l^{-1}(n-l)^{-1}$, placing more probability mass on the less
balanced splits close to $l=1$ and $l=n-1$. The AB model is a
specific case of the one-parametric (with parameter $\beta$)
family of beta-splitting models \cite{ALDOUSBetaSplitting1996}.
Statistical quantities computed from the AB ensemble (parameter
value $\beta=-1$) have been identified as giving a good fit to
real data
\cite{Aldous2001,BLUMWhichRandomProcess2006,Jones2011}.
The expected depth scales as $(\log n)^2$
\cite{ALDOUSBetaSplitting1996}. It is interesting to note that
the AB case $\beta=-1$ is precisely the critical point
separating two qualitatively distinct scaling behaviors in the
general beta-splitting model: standard logarithmic scaling for
$\beta>-1$, and power-law scaling $\langle d \rangle \sim
n^{-\beta-1}$ for $\beta<-1$ \cite{ALDOUSBetaSplitting1996}.

The AB model, beta-splitting, and other models \cite{Ford2006}
introduced to account for tree imbalance, however, assign
probabilities to tree shapes in a way which is not based on any
evolutionary mechanism. While they can statistically reproduce
features of the trees in the databases, this does not hint at
any biological explanation of these features, as
Ref.~\cite{BLUMWhichRandomProcess2006} remarks.

\section{The age model}

\subsection{The model and its depth}

We introduce the {\sl age model},
which describes the growth of a binary tree by iterative
stochastic addition of tips, one at each time step.
Each tip $i$ is assigned an age
$\tau_i(t)$ being the time that passed from the birth of the
tip, $t_i$, to the present time $t$, i.e.\ $\tau_i(t) = t -
t_i$. At time $t=0$ the tree consists of a single tip (the root), labeled
with the index $i=1$, representing an ancestral species.
The growth proceeds by iterating the
following three steps. (i) A tip $i$ is chosen with probability
$p_i(t)$ inversely proportional to a power of its age
\begin{equation}
\label{agedef}
p_i(t) = \frac{\tau_i^{-\alpha}}{c_\alpha(t)}~,
\end{equation}
where the normalization constant $c_\alpha(t)$ is chosen
such that probabilities from all tips sum up to 1; (ii) a new
branch $j$ is split from $i$ with creation time $t_j=t$ while
tip $i$ remains; (iii) time $t$ is increased by $\Delta t$ and
the process resumes at (i). Each branching represents a new
species evolutionarily splitting from the original one. 
This is coherent with a scenario of {\sl peripatric} speciation, in
which a small part of an ancestral population becomes isolated
and starts an independent evolution process, whereas the main
part of the population continues its previous dynamics.
We focus on $\Delta t=1$ first and discuss the case of $\Delta
t=1/n$ later, in section~\ref{sec:extensions}. There we also regard
a variation that treats both new descendants of the split branch
as new, starting at age zero. In terms of
biological evolution, the symmetric splitting
is allopatric speciation.

This defines a {\em family} of models parameterized by
$\alpha$. We note that $c_\alpha(n)$ (and $p_i(n)$) depends
only on the values of $n$ and $\alpha$ (and $i$). These
quantities are independent of the details of the previous
branching history, which is a stochastic process. The ERM
model, in which at each step one tip is uniformly
(independently of its age) chosen for speciation, is recovered
for $\alpha=0$. Negative $\alpha$ enhances branching
probability of the oldest tips, so that trees more balanced
than random ones are expected. We will see that
for sufficiently large positive $\alpha$ (in fact $\alpha\ge
1$) the excess branching probability given to the youngest tips
strongly breaks balance and modifies the ERM logarithmic
depth scaling.

Figure~\ref{fig:depth}(b) shows the dependence of mean depth on
tree size for the age model for several choices of the exponent
$\alpha$. At $\alpha=1.0$, we obtain $\langle d \rangle \sim (\log
n)^2$, both for time increments $\Delta t =1$ and $\Delta t
=1/n$ (see below). The parameters of the fitted curves agree
well between model and data (Figure~\ref{fig:depth}(a); see the
figure caption for details). For comparison, the
size-dependence of depth in the AB model is also shown.

\begin{figure}
\includegraphics[width=\columnwidth]{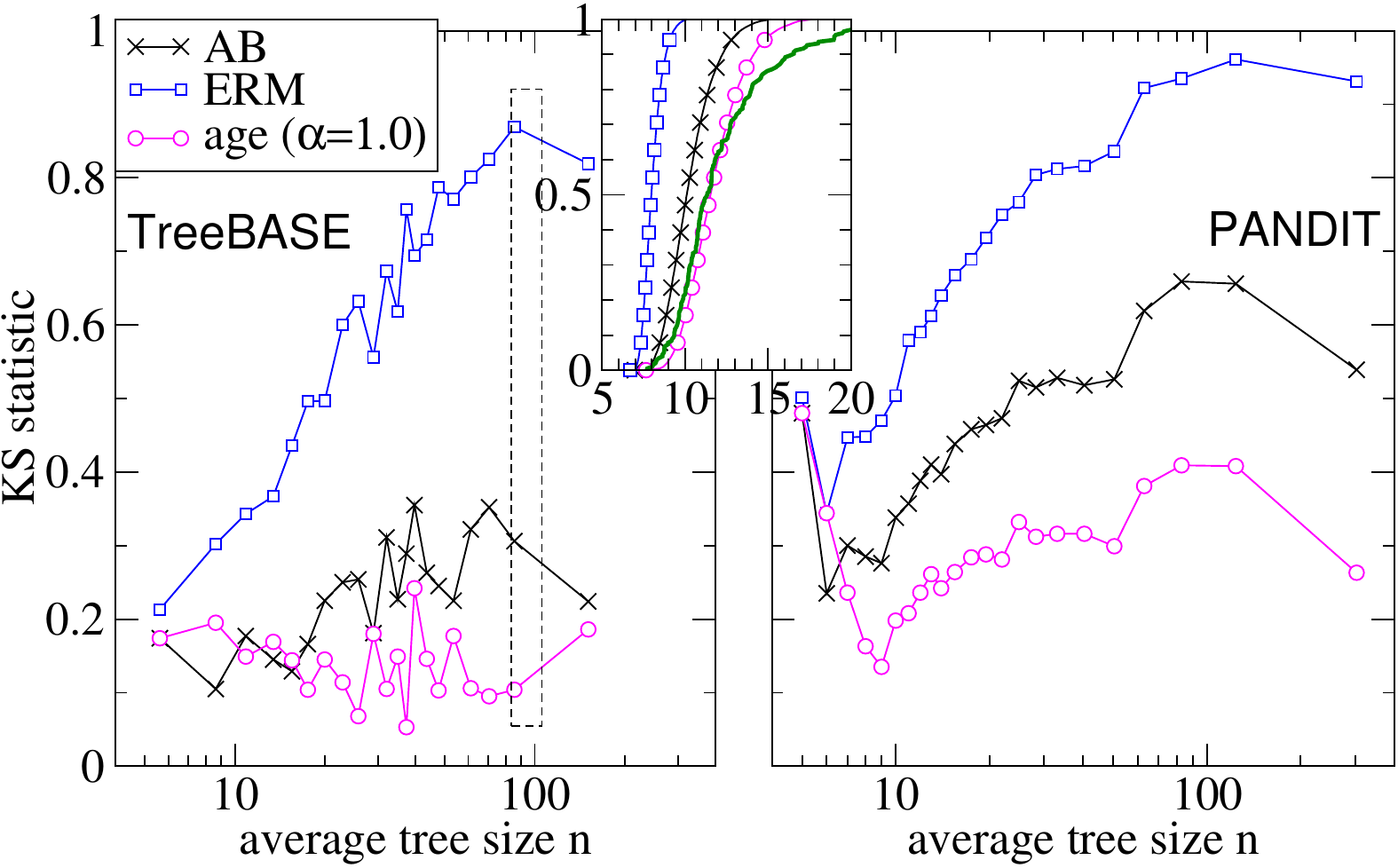}
\caption{ \label{fig:kolsmir} (Color online)
Comparison between data and models (AB, ERM and age) by distributions of depth.
The large panels
show the maximum deviation (Kolmogorov-Smirnov statistic) between the
cumulative distribution of depth in each model and the real trees.
Subsets $\mathcal{T}$ of the databases are
chosen the same as the bins in Figure~\ref{fig:depth}(a). See the main
text for further details. The inset shows, for one subset of TreeBase (trees of
size $76\le n \le 102$), the cumulative distributions of the real trees (thick
curve without symbols) and the three models, leading to the KS statistic values
marked by the dashed rectangle.
}
\end{figure}

Going beyond averages and considering also fluctuations, a
closer comparison between models and data is made in
Fig.~\ref{fig:kolsmir} by the Kolmogorov-Smirnov (KS)
statistic. For a set of real trees $\mathcal{T}$, the
cumulative depth distribution $q(d)$ is the fraction of trees
in $\mathcal{T}$ having depth less than $d$. For each tree in
$\mathcal{T}$, we generate with the rules of the model being
tested $100$ trees of the same size, obtaining a collection
$\mathcal{T}^\prime$ of $100\times|\mathcal{T}|$ model trees
having the same size distribution as $\mathcal{T}$ and a
cumulative depth distribution $q^\prime(d)$. The KS statistic
is the maximum deviation $\kappa = \max_{d \in \mathbb{R}}
|q(d) - q^\prime(d)|$ between data and model distributions,
with $\kappa=0$ if and only if the distributions are identical.
Except for the smallest trees ($n<20$), we find
(Fig.~\ref{fig:kolsmir}) that the depth distributions of the
real trees in both databases are systematically closer to the
age model with $\alpha=1.0$ than to the AB model.

\subsection{Analytic calculations for the age model and
finite-size corrections}

At time step $n$ (and taking $\Delta t=1$), the tree has $n$
tips with ages $\tau_1=n$, $\tau_2=n-1$, \ldots,
$\tau_i=n-i+1$,\ldots, $\tau_n=1$. Thus the normalization
constant is
\BE
c_\alpha (n) =  \sum_{i=1}^n
\frac{1}{\tau_i^{\alpha}}=\sum_{k=1}^n \frac{1}{k^\alpha} \ .
\label{cn}
\EE

The asymptotic behavior of $c_\alpha(n)$ for large $n$ is:
\BE
c_\alpha (n) \sim \left\{
\begin{array}{rl}
             \frac{n^{1-\alpha}}{1-\alpha} , & \text{if } \alpha<1   \\
             \log n , & \text{if } \alpha=1   \ \ , \textrm{ as } n\rightarrow\infty \ .\\
             \zeta(\alpha) , & \text{if } \alpha>1
\end{array}
\right. \label{cscaling}
\EE
$\zeta(\alpha)$ is Riemann's zeta
function, which is finite for $\alpha>1$.
The expected age of the tip chosen at time $n$ is
\BE
\overline{\tau(n)} = \frac{ \sum_{\tau=1}^n \tau^{1-\alpha}}{c_\alpha(n)}
\underset{n\rightarrow\infty}{\sim} \left\{
\begin{array}{rl}
             \frac{1-\alpha}{2-\alpha} n ,  &\text{if } \alpha<1   \\
             \frac{n}{\log n}            ,   &\text{if } \alpha=1  \\
             \frac{n^{2-\alpha}}{\zeta(\alpha)(2-\alpha)} ,  &\text{if } 1<\alpha<2 \ .        \\
             \frac{6}{\pi^2} \log n , &\text{if } \alpha=2         \\
             \frac{\zeta(\alpha-1)}{\zeta(\alpha)}   , & \text{if } 2<\alpha
\end{array}
\right. \label{tscaling}
\EE
This shows that the chosen age becomes progressively younger as
$\alpha$ increases. Older branches become less likely to branch
and imbalance is enhanced. Note that $\overline{\tau(n)} \sim
n/2$ for the ERM model ($\alpha=0$).

A heuristic argument to obtain $\langle d(n)\rangle$ uses
$\Delta n/\overline{\tau(n)}$ as an estimate of the mean number
of branching events in a time interval of length $\Delta n$
centered at $n$. Thus we can count the mean number of branching
events as a function of $n$ by the integral:
\BE
\langle d(n)\rangle \approx \int_{1}^n
dn'\frac{1}{\overline{\tau(n')}}
\underset{n\rightarrow\infty}{\sim} \left\{
\begin{array}{rl}
             \log n ,                               &\text{if } \alpha<1   \\
             (\log n)^2            ,                  &\text{if } \alpha=1  \\
             n^{\alpha-1} ,                         &\text{if } 1<\alpha<2    \ .    \\
             {\rm li}(n) \sim \frac{n}{\log n}  ,   &\text{if } \alpha=2         \\
             n   ,                                  &\text{if } 2<\alpha
\end{array}
\right.
 \label{dscaling}
\EE
Prefactors have not been included since our crude argument is
not expected to give them exactly. ${\rm li}(n)$ is the
logarithmic integral function.

\begin{figure*}
\includegraphics[width=0.97\textwidth]{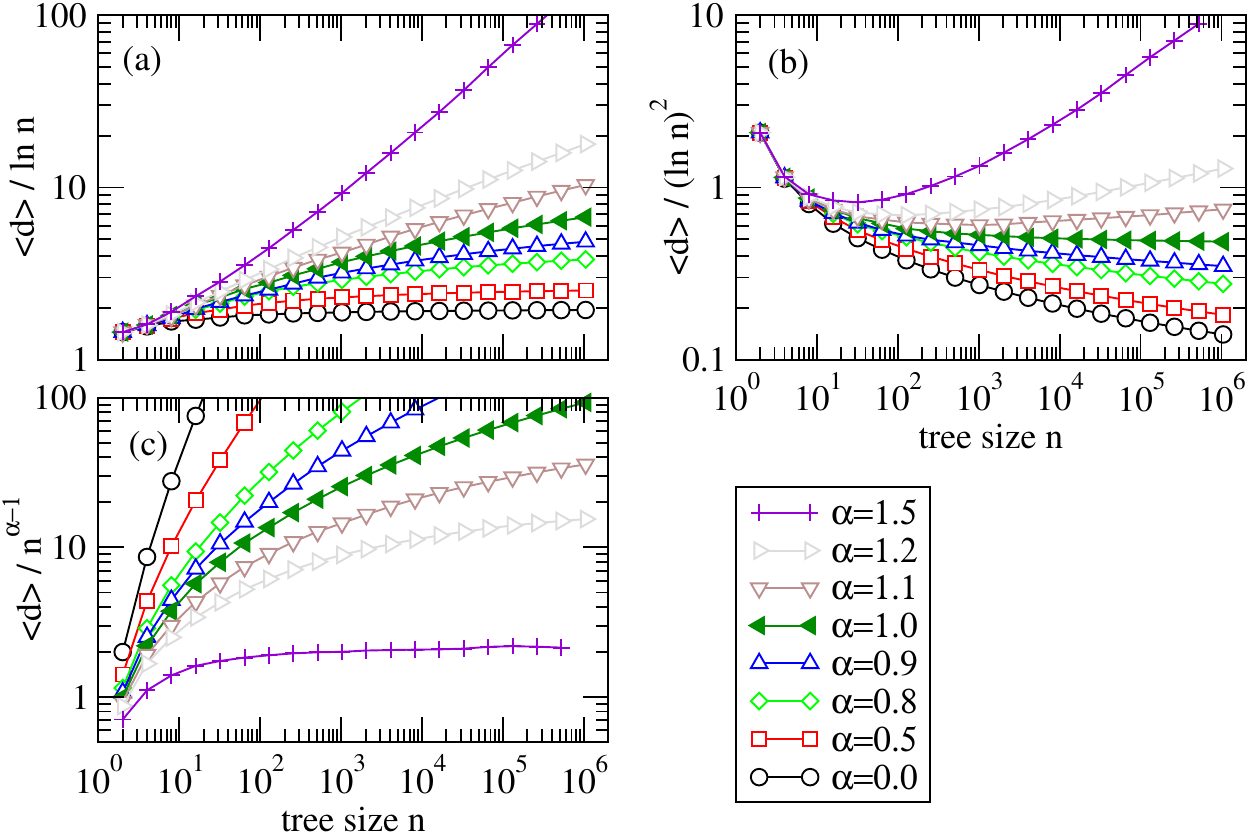}
\caption{\label{fig:finitesize} (Color online) Analysis of finite size
corrections. We plot the average depth of trees from the
age model for various parameter values $\alpha$.
Different panels use different rescaling of depth, see the
y-axis labels. If finite size corrections from the scaling of
Equation (\ref{dscaling}) were absent, the three lower curves
($\alpha<1.0$) would be flat in panel (a). In panel (b) this
would be the case exactly for $\alpha=1.0$ (filled triangles).
In panel (c), the three curves for $\alpha>1.0$ would be flat.
Each data point is an average over 100 independent trees.}
\end{figure*}

Equation~(\ref{dscaling}) gives the $n$-dependence of the
depth $\langle d  \rangle$ in leading order for large tree size
$n$. Finite size corrections to scaling, found numerically, are
significant only close to the transition.
Figure~\ref{fig:finitesize} shows finite size corrections to
this leading order. As usual, the corrections become
large close to a transition point, in this case close to
$\alpha=1$ (Figure~\ref{fig:finitesize}(a)). Likewise,
$\alpha$ above but close to $1.0$ leads to large corrections
from the scaling with $n^{\alpha-1}$
(Figure~\ref{fig:finitesize}(c)). For $\alpha=1.0$, the
function $\langle d \rangle / (\ln n)^2$ is only weakly
dependent on $n$, falling monotonically from 0.85 to 0.48 over
5 orders of magnitude in $n$ (filled triangles in
Figure~\ref{fig:finitesize}(b)).

The predicted behavior (\ref{dscaling}) is consistent with our
numerical findings. Thus the age model with
$\alpha=1$ leads to the asymptotic square-logarithmic scaling
that appears to describe the real phylogenies rather well.
Interestingly, this particular scaling appears at the critical
transition between purely logarithmic and power-law scaling, in
much the same way as for the beta-splitting model
\cite{ALDOUSBetaSplitting1996}. This may indicate a kind of
universality in tree-shape transitions, and associated
universality classes. Another transition, to the comb tree
scaling, occurs at $\alpha=2$.

\subsection{Extensions} \label{sec:extensions}

The definition of the age model so far describes {\sl peripatric}
speciation. A more
symmetric {\sl allopatric} speciation mechanism would imply a
more similar role for the species arising in a branching event,
for example a resetting to zero of the age of the two species
emerging from the branching, so that both are considered to be
{\sl new} and not just one of them. The analytic evaluations
become more delicate, since $c_\alpha(t)$ is now a random
variable, but heuristics confirms that the
asymptotic scaling of the expected age $\overline{\tau(n)}$ and
depth scaling $\langle d(n)\rangle$ in the symmetric model is
the same as for the corresponding asymmetric one given by
Eqs.~(\ref{tscaling}) and (\ref{dscaling}). Numerically we find
the mean depth obtained from the allopatric version to coincide
with that of the original peripatric version of the model at
$\alpha =1$: Relative deviations between $\langle d \rangle$
estimates are below $1 \%$ and become smaller for growing $n$.

Another important extension corresponds to the case in which
the {\sl age} of the tips is not measured in number of
speciation events $\tau$, but in a different but related time
unit. It is biologically reasonable (as assumed also in the
Yule model) that speciation rate is proportional to the number
of species present, so that the instants of times assigned to
successive speciation events $t_n$ and $t_{n+1}$ are related to
the numbering of speciation events $n$ and $n+1$ by $\Delta
t\equiv t_{n+1}-t_n=1/n$. This implies $t_n \sim \log n$ at
large $n$. This new time-age $a_n(\tau)$ of a tip that has an
event-age $\tau$ is thus $a_n(\tau)=t_n-t_{n-\tau}\sim \log n -
\log(n-\tau)=-\log(1-\tau/n)$ for large $n$. For a version of
the age model with speciation probabilities proportional to
$1/a_n^\alpha$, we can recalculate the expected value of the
event-age $\overline{\tau(n)}$ chosen at instant $n$ or $t_n$:
\BE
\overline{\tau(n)}=\frac{\sum_{\tau=1}^n \tau
a_n(\tau)^{-\alpha}}{\sum_{\tau=1}^n  a_n(\tau)^{-\alpha}} \ .
\EE

To further analyze this expression, we approximate the sums by
integrals, and introduce the change of variable $s=\tau/n$.
After this it is clear that the integrals for large $n$ are
dominated by the singularities arising as $s\rightarrow 0$ (say
within the interval $s\in [1/n,\epsilon]$, with $\epsilon$
small), which allow us to use the small $s$ expansion:
$a_n(\tau)\approx -\log(1-s) \approx s$:
\BE
\overline{\tau(n)} \approx \frac{\int_{\frac{1}{n}}^\epsilon n
s \left( s^{-\alpha}+ \ldots
\right)ds}{\int_{\frac{1}{n}}^\epsilon \left(s^{-\alpha}+
\ldots \right)ds } \ . \label{meantauextended}
\EE
Now, these integrals become identical to the ones corresponding
to the asymptotic evaluation of the sums in the original age
model, so the asymptotic behavior of the depth will be the
same. This points out, once again, that the important
ingredient needed to alter the standard logarithmic ERM depth
scaling is the excess of branching probability assigned to
young branches (small $\tau$) by the $\tau^{-\alpha}$ factor.
Continuous-time branching processes in which different branches
split independently with some waiting time distribution of the
renewal type \cite{Harris1963} can be also considered. They
would have instantaneous branching rates decaying as $t^{-1}$
at long times when the waiting-time density decays as a power
law at these long times. But as Eq.~(\ref{meantauextended})
reveals this long-time behavior is not the relevant one, but
the presence of a short-time singularity. Singularity with the
required strength can not be achieved with normalizable
continuous-time waiting-time densities. This is why we have
always found the normal depth scaling as $\log n$ in
simulations of this type of processes, even when the waiting
time distribution had fat tails. Another ingredient in
the age model is that the normalization constant in the
denominator of Eq.~(\ref{agedef}) involves the age of all
branches. This provides an interaction among all branches,
which is absent in models of independent branching.

\section{Discussion}

Motivated by observations of anomalous scaling in evolutionary
trees
\cite{BLUMWhichRandomProcess2006,Herrada2008,Jones2011,Herrada2011,Pompei2012},
the proposed {\em age model} introduces time-correlations
and branch interactions such that a variety
of depth scalings can be reached beyond the standard
logarithmic one. Remarkably, for the case $\alpha=1$,
corresponding to the critical point between two qualitatively
different {\sl phases}, the model agrees with observed
phylogenetic trees better than previous models. In addition, it
describes the tree generation process assuming that lineages
which have not speciated for a long time display in the future
a still more reduced speciation rate. This kind of phenomenon
is discussed in the framework of evolution and heritability of
evolvability and robustness in the biological literature
\cite{Masel2009}, and of {\sl phenotypic entrapment} in
genotype networks \cite{Manrubia2013}. In a broader
picture, dynamics on possibly rugged fitness landscapes
\cite{Sibani:1995} provides evolution at the microscopic level.
It could serve as a mechanistic explanation of the assumptions
of the age-dependent model.

Future work should consider the inclusion of extinction
processes into the model. This is a realistic element that
would open the possibility of an additional critical behavior
(the transition between growth and extinction) known to
alter tree topology \cite{Harris1963,Newman1999,DeLosRios2001,HernandezGarcia2010,Becker2014}.
Why evolution should be poised at the critical point deserves further
investigation \cite{Bak:1993,Sole:1996}. Analyses and comparison of branch
length distributions are worth pursuing. Although branch length data of phylogenies are
not as reliable as their topological structure
\cite{BarracloughNeePhylogeneticsAndSpeciation2001},
improvements are rapidly accumulating (see e.g.
Ref.~\cite{Venditti2010}).

\acknowledgments
We thank Alejandro Herrada, Stephan Steigele and Joan Pons for
valuable discussions regarding biological evolution. This work
has been supported by MINECO (Spain) and FEDER (EU) through
projects INTENSE@COSYP (FIS2012-30634) and MODASS
(FIS2011-24785), and by VolkswagenStiftung through contract
I~/~82~719.

\bibliography{AgeModel}

\end{document}